\documentstyle[11pt]{article}
\documentstyle{usepictex}
\title{Numerical Solution of the Schroedinger Equation
using a Quantum Lattice Boltzmann Equation}
\author{S. Succi
\thanks {IBM European Center for Scientific and Engineering Computing,
via Oceano Pacifico 171, 00144, Roma, Italy} }

\begin{document}
\maketitle
\bibliographystyle{plain}

\section{Abstract}

The quantum Lattice Boltzmann equation (QLBe), a new variant of the
lattice Boltzmann equation, specifically designed to describe
non relativistic quantum motion, is validated for the
case of a free-particle in (1+1) space-time dimensions.
The proper parameter regime under which the method needs
to be operated in order to reproduce faithful non-relativistic
quantum motion is also discussed.

\section{Introduction}

The Dirac equation is the most general equation
describing single particle motion in compliance with the
two fundamental columns of modern physics; quantum theory and special
relativity.
Relativistic effects rise with the ratio $\beta$ of particle
$v$ to light speed $c$, while quantum effects are driven by
non-vanishing values of the diffusion coefficient $D=\hbar/2m$, $m$ being
the particle rest mass and $\hbar$ the Planck's constant.

Thus, in the plane $(\beta,D)$, one can identify four qualitative
regions associated with relativistic quantum mechanics
($\beta \sim 1$, $D \sim 1$ ), described by the Dirac equation,
non-relativistic quantum mechanics
($\beta \ll 1$, $D \sim 1$ ), described by the Schroedinger equation,
relativistic classical mechanics
($\beta \sim 1$, $D \ll 1$ ), described by the special-relativity and
non-relativistic classical mechanics
($\beta \ll 1$, $D \ll 1$ ), described by the Newton equation,
Here, $D$ is normalized to $l^2/\tau$, where $l$ and $\tau$ are typical
spatial and temporal scales over which the dynamics is observed.

In a recent paper, a new procedure to solve quantum mechanical
problems using numerical techniques mutuated from discrete kinetic
theory has been proposed /1/.
This procedure builds on a formal analogy between the
Dirac equation and a special discrete kinetic equation known as
Lattice Boltzmann equation (LBe) (for a review see /2/).
In particular, it was shown that by a proper resort to operator
splitting methods, the Dirac equation can be integrated as a
sequence of three one-dimensional LBe's evolving complex-valued
distribution functions.

It is the purpose of this letter to investigate the ability of the
numerical procedure outlined in /1/ to properly describe the
non-relativistic quantum motion starting from a relativistic equation
such as the Quantum Lbe (QLBe).

\section{The Quantum LBe}

For the sake of simplicity, we shall confine our discussion to
one-dimensional motion (1+1) of a free quantum relativistic particle.
In the Majorana representation /3/, the Dirac equation reads as
\begin{eqnarray*}
\partial _t u - c \partial _z u = -\omega_c d \\
\partial _t d + c \partial _z d = +\omega_c u
\end{eqnarray}
where $u$ and $d$ represent a pair of complex bispinors and
$\omega_c=m_0 c^2/\hbar$ is the Compton frequency.
According to the procedure proposed in /1/, the system of equations (1) is
discretized as
follows
\begin{eqnarray*}
\hat{u} - u = a u + b d \\
\hat{d} - d = a d - b u
\end{eqnarray}
where $u=u(z,t), d=d(z,t)$  and
$\hat{d} = d(z-1,t+1)$,  $\hat{u} = u(z+1,t+1)$.

The scattering elements are given by
$a= (1-m^2/4)/(1+m^2/4)$ and $b= m/(1+m^2/4))$ where $m$
represents the dimensionless Compton frequency in lattice units
$\Delta x = \Delta t = c = 1$.

The equation (2) represent a Cranc-Nicolson time-marching scheme
which is known to preserve unitarity ($a^2 + b^2 = 1$) all along
the time evolution.

As noted in /1/, the eq.(2) can be interpreted as a discrete Boltzmann equation
for a pair of complex wavefunctions, $u$ and $d$, (counter)streaming along the
$z$ axis
and undergoing collisions according to scattering matrix on the
right-hand-side.

In the limit $m \rightarrow 0$, the eq.(1) describe dispersionless
free-propagation of two counterpropagating light pulses.
As $m$ takes non zero values, 'interspinor collisions' take place
which produce both drag and diffusion effects on the wavefunction.
Drag results in subluminal propagation speed ($v < c$)
of the wavepacket whereas diffusion is responsible for its
spatial delocalization.

It is worth noticing how, in this view, both quantum
and relativistic effects can be tracked back to a simple common
origin: a non-zero particle mass.
This is somehow suggestive of "hidden-variables" theory, whereby the
distribution
function plays the role of a coarse-grained variable resulting from the
collective motion of a cloud of unobservable "automata" all propagating
synchronously in lock-step mode at light-speed and undergoing collisions
according to the r.h.s of equation (1).
Leaving aside this fascinating and controversial subject, we turn to the issue
of the numerical validationof the QLBe.

\section{Numerical test: Free propagation}

The non-relativistic limit is most conveniently discussed by inspecting the
dispersion
relation associated with the equation (1).
This reads

\begin{eqnarray*}
\omega_{\pm}  = \pm \sqrt {k^2 + m'^2}
\end{eqnarray}

where $m'= (1/ \Delta t ) arctan(\frac{\mu}{1-\mu^2/2})$
is an effective mass incorporating the effects of lattice discreteness
($\mu = m \Delta t $).
In the limit $\Delta t \rightarrow 0$, $m'$ reduces to $m$ as
imposed by consistency requirements.
This sets an upper bound to $m$.
This bound is however rather 'soft', because, due to the
shape of the arctan function, the deviations of $m'$ from $m$ remain
quite small except for $m$ pretty close to one.

As a result, in the following we shall always refer to m, with the tacit
assumption $m \ll 1$.

The Schroedinger equation follows from the Dirac equation in the
limit of long-wavelengths, low frequencies:
\begin{eqnarray*}
k^2 \ll m^2
\end{eqnarray}
or equivalently, in lattice units

\begin{eqnarray*}
\beta = k/m \ll 1
\end{eqnarray}
In order to test the validity of our scheme in this regime, we
evolve the following pair of minimum uncertainity wavepackets

\begin{eqnarray*}
\Psi_{u,d} = \frac{1}{\sqrt {2 \pi T_0}} exp -\frac{(z-z_0)^2}{2T_0}
exp \pm i \beta_0 (z-z_{0}/2)
\end{eqnarray}

This expression represents two wavepackets centered about $z_0$ with an
initial spread $\Delta_0 = \sqrt T_0$ propagating at speed
$\beta_0$ along the $+z$ (particle) and $-z$ (anti-particle) axes respectively.

Following a standard procedure, we introduce a pair of
transformed wavefunctions defined as follows:

\begin{eqnarray*}
\phi _{\pm} = \frac{1}{\sqrt 2} (\phi_u \pm i \phi_d) exp (-imt)
\end{eqnarray}

which represent the slow (+) hydrodynamic and fast (-) non-hydrodynamic
modes. Note, that these modes combine a weighted mix of particle ($u$)
and antiparticle ($d$) states.

It is a simple matter to show that
the fast mode amplitude is order $O(\beta)$ with respect to the hydrodynamic
mode
and its oscillation frequency is $O(\beta^{-1})$.
As a result, in the adiabatic limit $\beta \rightarrow 0$, the fast mode
decouples from the
system dynamics which is consequently
 dominated by the slow mode.
It is precisely under these conditions that non-relativistic quantum motion is
expected to
emerge out of a relativistic equation such as eq.(1).

To check the ability of QLBe to reproduce this dynamical decoupling, we
monitored the main representative parameters
associated with the wave motion, i.e. the mean displacement
$\zeta = < z >$ and the mean spread $\Delta = (<z-\zeta>^2)^{1/2}$.

As is well known /4/, these quantities evolve accordingly to the following
exact equation

\begin{equation}
\zeta(t) = z_0 + \beta t
\end{equation}
and
\begin{equation}
\Delta(t) = \sqrt { \Delta_0^2 + \frac{t^2}{4m^2\Delta_0^2} }
\end{equation}
\\
In figure 1 we show $\zeta - z_0$ and $\Delta$ as a function of time
for the following choice of parameters
$m=0.1, \; \Delta_0=50$, $z_0 = 1024$, on a $2048$ mesh.

The suffix (+) denotes averaging over $\rho_+ = |\phi_+^2|$.
{}From this figure, one can appreciate a remarkable match of the (+)
quantities with the analytical results (the short solid segments indicate
the analytical slope derived from eqs(7-8) in the limit $t \gg 1$).

In Fig.2 we show the probability distribution function $\rho_+$
as a function of space for three different times.
The typical drift-diffusion behaviour is clearly visible.

A series of runs at different values of $\beta$ invariably showed that
indeed the probability density of the
non-hydrodynamic mode is order $\beta^2$ with respect to the hydrodynamic one.
These results prove that the adiabatic limit of eq.(1) correctly
describes non-relativistic Schroedinger dynamics.
\\
An interesting question arises on to whether the present scheme is also able
to describe the onset of relativistic motion.
This point has been investigated by systematically varying the particle
speed $\beta_0$ in the range (0,1).
Having defined $\beta_0 = p/m$, $m$ being the {\bf rest} mass, relativistic
motion
is represented by the following equation

\begin{equation}
\frac{d \zeta(t)}{dt} = \beta
\end{equation}
where
\begin{equation}
\beta = \beta_0 (1 + \beta_0^2)^{-1/2}
\end{equation}

The equation (11) reflects the inclusion of the velocity dependence of the
particle
mass $m(\beta)= m/(1 -\beta^2)^{1/2}$ in the relativistic equation of motion
$dp/dt \equiv d m(v)v/dt=0$.
The same expression can obviously be derived by computing the group velocity
$\partial \omega / \partial k$ from the dispersion relation eq.(3).

The results are presented in Fig.s 3-4 where the asymptotic values of
$\dot \zeta$ and $\dot \Delta$ are shown as a function of $\beta_0$.
Figure 3 shows that $\zeta$ and $\zeta_+$ are pretty close to the
exact results up to $\beta \sim 0.3$, which is in good match with the
theoretical expectations.

For higher beta's, relativistic effects become manifest
in the form of a visible departure of $\beta$ from $\beta_0$ as
expressed by the equation (11).

Note however, that only $\zeta_+$ keeps following closely the correct solution,
while
$\zeta$ displays a considerable excess of 'slowing down'.
This is perfectly in line with the fact that all physical quantities pertaining
to
the hydrodynamic mode should be weighted with the probability
density $\rho_+$ rather than with the total density $\rho=\rho_+ + \rho_-$.

At low beta this doesn't make any difference because $\rho_+ \sim \rho$; at
high beta however, the fast mode 'grows up' and this difference becomes quite
visible.

At high beta, the hydrodynamic mode looses its connotation
of a slowly drifting and diffusing single-bumped wavefunction, and takes the
form of two
well distinct counterpropagating bumps.
Under these conditions, the equation (9) for the mean spread clearly
breaks down, as witnessed by figure 4, whereas the mean
displacement is still given by the eq.(8) provided the relativistic
expression for beta is accounted for.

At this point is worth mentioning that the quality of the results is affected
not only by the value of $m$ but also by the choice
of the initial spread $\Delta_0$. This parameter must be kept sufficiently
large
so as to prevent high wavenumbers from violating the adiabaticity conditions.

Note in fact, that the range of wavenumbers active during wave motion is
approximately given by

\begin{equation}
| k | \sim  | k_0  + 1/ {\Delta(t)} | < | k_0 + 1/{\Delta_0} |
\end{equation}

This relation shows that adiabaticity can be violated whenever $\Delta_0 <
1/k_0 $ even
though $k_0$ lies well within the non-relativistic range.

This inequality is indeed fulfilled for the present simulations in which the
value $\Delta_0 = 50$ has been chosen.

Further simulations at smaller values of $\Delta_0$ clearly indicated that
trustful non-relativistic motion can be tracked with the quantum LBe
provided the smallest wavelengths are well beyond a minimal value
of the order of about $1/m$ lattice units.

We are now in a position to clearly identify the parameter regime under which
QLBe provides
a faithful numerical solution of Schroedinger dynamics.
The particle mass cannot be too large for reasons of numerical accuracy;
it cannot be too small either because otherwise adiabatic relaxation takes so
long
time intervals and so large spatial regions to set in, that numerical
efficiency
would be completely spoiled.

A reasonable choice is $m$ between $0.1$ and $0.5$.
Finally, the initial conditions shall also be chosen in such  a way as to
preserve
compatibility with the adiabatic assumption.

\section{Conclusion}

The present results indicate the the Quantum LBe is indeed able to reproduce
simple
non-relativistic quantum mechanical motion.
The onset of relativistic single-particle motion seems also well reproduced.
Extension to more complex situations (higher dimensionality and interacting
systems) will
be subject of future research.

\newpage

\section{Figure captions}
\begin{itemize}
\item Figure 1:

Mean (curves 1,1+) and variance (2,2+) of the wavefunction as a function of
time
for $\beta_0 = 0.2$ and $m=0.1$.
The suffix (+) indicates average over the density of the hydrodynamic mode.

\item Figure 2

Probability distribution $\rho_+$ of the hydrodynamic mode at three different
times $t=0,1000,2000$, for the same parameters as Fig. 1.
A typical drift-diffusion behaviour is clearly visible.

\item Figure 3

Time derivative of the mean displacement of the hydrodynamic mode (big crosses)
and total wavefunction (squares) respectively as a function of $\beta_0$.
The solid curve represents $\beta$ as a function of $\beta_0$ as egiven by
the equation (11), while the straight line is $\beta=\beta_0$.
The mass is set at $m=0.1$, and the initial spread is $\Delta_0=50$.

\item Figure 4

The same as figure 3 for the mean spreading.

\end{itemize}

\end{document}